\documentclass[fleqn,usenatbib]{mnras}

\usepackage{newtxtext,newtxmath}

\usepackage[T1]{fontenc}

\DeclareRobustCommand{\VAN}[3]{#2}
\let\VANthebibliography\thebibliography
\def\thebibliography{\DeclareRobustCommand{\VAN}[3]{##3}\VANthebibliography}

\usepackage{graphicx}	
\usepackage{amsmath}	

\newcommand\cm{{\rm\thinspace cm}}

\newcommand\K{{\rm\thinspace K}}

\newcommand\Msun{\hbox{$\rm\thinspace M_{\odot}$}}

\newcommand\pc{{\rm\thinspace pc}}

\newcommand\yr{{\rm\thinspace yr}}

\newcommand\Msunpyr{\hbox{$\Msun\yr^{-1}\,$}}

\newcommand\pcmsq{\hbox{$\cm^{-2}\,$}}
\newcommand\pcmK{\hbox{$\cm^{-3}\K$}}

\title[Early low-mass star formation]{Consequences of a low-mass, high-pressure, star formation mode in early galaxies}

\author[A. C. Fabian et al.]{
A. C. Fabian,$^{1}$\thanks{E-mail: acf@ast.cam.ac.uk },  J.S. Sanders$^{2}$, G.J. Ferland$^{3}$,  B.R. McNamara$^{4}$, C. Pinto$^{5}$ and S.A. Walker$^{6}$
\\
$^{1}$Institute of Astronomy, University of Cambridge, Madingley Road, Cambridge CB3 0HA, UK\\
$^{2} $Max-Planck-Institut fur extraterrestrische Physik, Giessenbachstrasse 1, 85748 Garching, Germany\\
$^{3} $Department of Physics, University of Kentucky, Lexington KY 40506, USA\\
$^{4} $Department of Physics and Astronomy, University of Waterloo, 200 University Avenue West, Waterloo, ON N2L 3G1, Canada\\
$^{5} $INAF-IASF Palermo, Via U. La Malfa 153, I-90146 Palermo, Italy\\
$^6$ Department of Physics and Astronomy, The University of Alabama in Huntsville, Huntsville, AL 35899, USA\\ }
\date{Accepted XXX. Received YYY; in original form ZZZ}

\pubyear{2024}

\begin{document}
\label{firstpage}
\pagerange{\pageref{firstpage}--\pageref{lastpage}}
\maketitle

\begin{abstract}
High resolution X-ray spectra reveal hidden cooling flows depositing cold gas at the centres of massive nearby early-type galaxies with little sign of normal star formation. Optical observations are revealing that a bottom-heavy Initial Mass Function is common within the inner kpc of similar galaxies. We revive the possibility that a low-mass star formation mode is  operating  due to the high thermal pressure in the cooling flow, thus explaining the accumulation of low-mass stars. We  further explore whether such a mode operated in early, high-redshift galaxies and has sporadically continued to the present day. The idea links observed distant galaxies with black holes which are ultramassive for their stellar mass, nearby red nuggets and massive early-type galaxies. Nearby elliptical galaxies may be red but they are not dead. 
\end{abstract}

\begin{keywords}
galaxies: clusters: intracluster medium: black holes
\end{keywords}



\section{Introduction}
We have found that cooling flows are operating in many massive early-type galaxies HCFI, II, III \citep{Fabian22, Fabian23a, Fabian2023b}. The rapidly cooling, soft X-ray part of these flows is mostly hidden from direct view by photoelectric absorption in surrounding cold gas, much of which may be the product of the cooling process. Such Hidden Cooling Flows (HCF) have mass cooling rates (HCR) of $1-3\Msunpyr$ in typical massive Early-Type Galaxies (ETG) rising to $10-20\Msunpyr$ in the Brightest Group Galaxies (BCG) to $20-100{\rm s}\Msunpyr$ in Brightest Cluster Galaxies (BCGs). Some exceptional clusters have even higher rates of $\dot M\sim 1000\Msunpyr$. 

Jetted AGN Feedback (FB) probably reduces the total inflow rates by a factor 2 to 5, mainly affecting the outer parts of the flow beyond the central kpc. It is difficult to envisage how it could reduce the rates by a further factor of 10 or more which would be required if there is no absorption affecting the soft X-ray emission.   The new X-ray absorption-corrected HCR are the mass cooling rates after AGN FB is included, but nevertheless far exceed the  Star Formation Rate (SFR) in most early type galaxies and BCG, as discussed in Section 2.   The mismatch brings into focus the old Cooling Flow Problem of what happens to the cooled gas? Cool cores are found in about half of all clusters and groups of galaxies and are seen out to beyond redshift $z=1$ e.g. \citep{Masterson2023}. They must therefore be long-lived with ages of at least 5 Gyr, implying the accumulation of at least $5\times 10^9\Msun$ of cooled gas in typical elliptical galaxies, $5\times 10^{10}\Msun$ in rich groups and $1-10\times 10^{11}\Msun$ in BCGs. 

Several possible explanations are put forward in HCF1. They are cooling into a) undetectably cold molecular clouds, b) low mass star formation from such clouds and c) cycling of cooled gas to outer (kpc or more) parts of the cool core. Here we concentrate on the consequences of low-mass star formation. The possibility of a mode of low-mass star formation  in elliptical galaxies is not new and was first raised by \citep{Jura1977}. It was briefly mentioned in the cooling flow paper by \citep{Cowie1977} and discussed in detail by \citep{Fabian1982,Sarazin1983, Ferland1994},  and more recently by \citep{Sharda2022, tanvir2022, Grudic2023}). 

Dust and reddening may play some role here in reducing the SFR but in general we can state that whatever forms from the cooled gas, most of it is not normal stars. We propose here that it is low mass stars, brown dwarfs and lower mass objects, i.e. a mode of low-mass star formation, associated with the high pressure environment at the centres of HCF. It essentially forms Baryonic Dark Matter (BDM). 

As discussed in HCFIII, some brown dwarfs may be swallowed whole, without emitting radiation, into the central black hole if its mass $M_{\rm BH}>10^7\Msun$, The low-mass star formation mode dominates in the HCF of local early-type galaxies and we now explore its role in early galaxies at moderate to high redshifts. We touched on this last aspect briefly in HCFIII and in \citep{Thomas1990, Nulsen1995}, but explore it further here. There will have been a cooling flow phase in the formation of most galaxies following the initial gas collapse and starburst phase, resulting in significant BDM components. 

The purpose of this paper is to draw attention to the possibility that  a low-mass, high-pressure mode of star formation operates throughout  much of the lifetime  of a  massive galaxy. They continue to have atmospheres of hot gas fuelling a central cooling flow,  hidden from direct view at the present time by absorption from the cooled gas. Even when they otherwise look quiescent they may have low-mass star formation at their centre, as well as a silently growing black hole. 

\begin{figure}
	\includegraphics[width=0.53\textwidth]{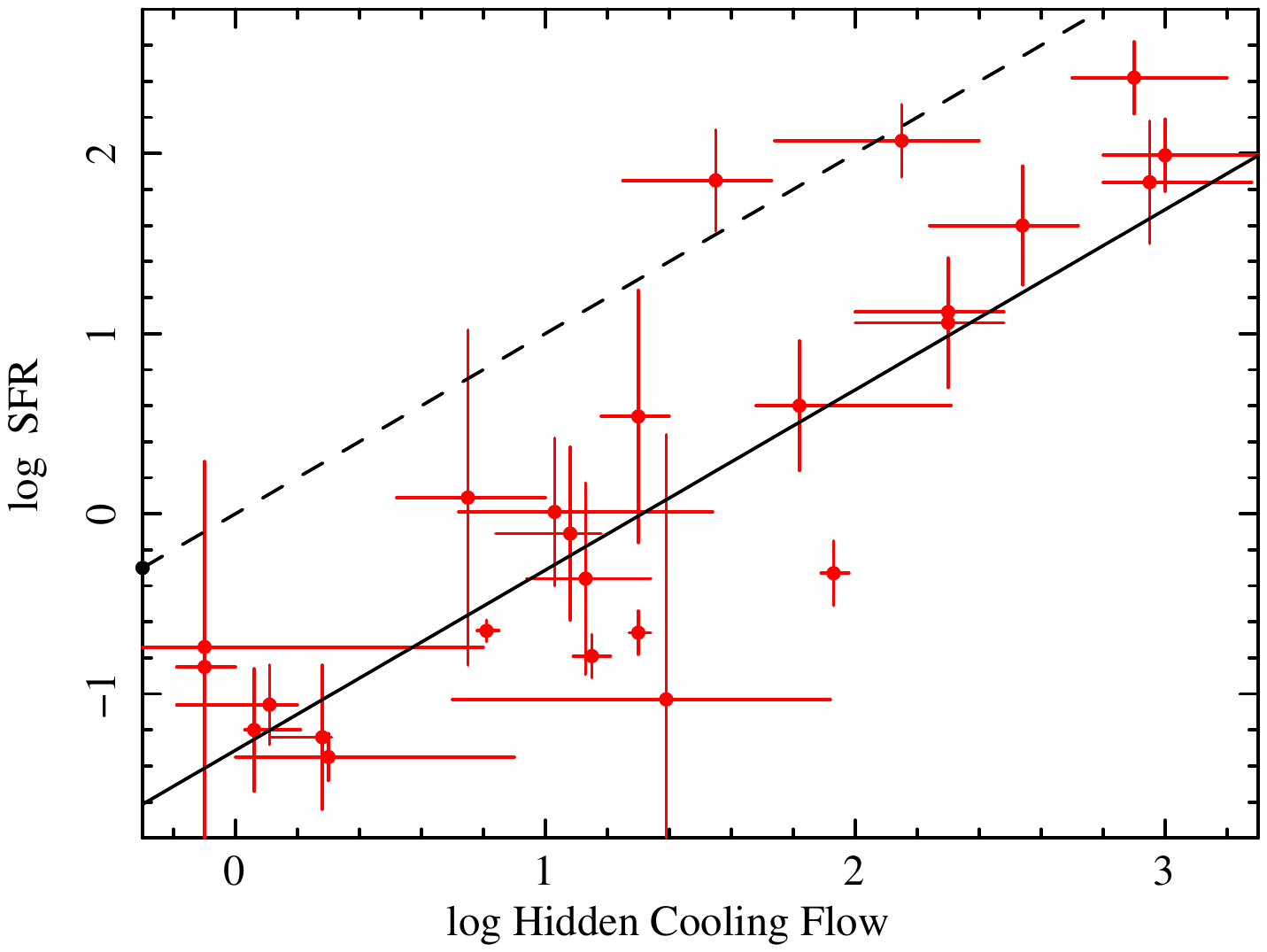}
 
    \caption{The Star Formation Rate (SFR) from \citet{McDonald2018} plotted against the Hidden Cooling Rate (HCR) from HCFI-IV.   The objects from left to right are: NGC1600, M87, NGC4472, NGC5846, M84, A2199, A262, Sersic 159, A3581, A2052, Centaurus, A1795, NGC5044, A85, Perseus, A2597, 2A0335+096, A1835, A1664, A1068, Cyg A, RXJ1532, Zw3146, MACS1931. The upper dashed line shows where the SFR and HCR are equal. The lower solid line is ${\rm SFR}={\rm HCR}/20$. }
    \label{fig:example_figure}
\end{figure}

\section{The Normal Star Formation Rate and the Hidden Cooling Rate}

We have plotted the Normal Star Formation Rates (SFR) against the Hidden Cooling Flow Rates (HCR) in Fig. 1. This follows the work of \cite{McDonald2018} who plot SFR against cooling rates derived from imaging data with no absorption considered, we denote the imaging rates as ICR. They find that for objects with ${\rm ICR}>30\Msunpyr$, ${\rm SFR}=(0.014\pm 0.004) \times {\rm ICR}.$ We use the SFR values from \cite{McDonald2018} which have been gleaned from the literature and are referenced there. The  SFR values are derived from emission lines, UV continuum or reradiated dust emission in the far infrared (FIR), so are dominated by emission from O and B stars within the host galaxy. 

HCR are generally smaller  than ICR due to AGN FB. Our values are  from HCF I, II and  III \citep{Fabian22,Fabian23a,Fabian2023b} and a further paper (HCFIV) in preparation. They have been measured using data from the XMM Reflection Grating Spectrometers (RGS) operating over the wavelength range of 8--22\AA\ and  approximately cover the inner arcmin of the target object,  where the emission lines from gas cooling below $10^7\K$ emerge. They yield the  mass cooling rates of gas cooling through the soft X-ray band. The spectra require photoelectric absorption of cold gas which we implement in a multilayer model in which hot and cold gas are interleaved, hence the name hidden cooling flow. Hubble Space Telescope imaging of BCGs often shows dusty clouds and patches at their centres (see HCFII \cite{Fabian23a}). 

If the HCR and SFR rates were equal they would lie along the upper dashed line. Instead we find them scattered along the best-fitting lower line which is where ${\rm SFR}=0.05 \times {\rm HCR}$. Clearly, most of the cooled gas does not end up in normal star formation.  The 2 objects to the upper right which do lie on the equal line are the Perseus cluster and A1835. Both are known for obvious and widespread normal star formation, much of which is extended well beyond the central kpc where most of the final cooling of the HCR occurs. The normal star formation in these objects does not concern us here and will be ignored for now. 

A good example of a well-studied nearby galaxy with a hidden cooling flow of $12\pm2\Msunpyr$ (HCFI) but very little star formation is NGC4696, the BCG in the Centaurus cluster (A3526) at a distance of 45 Mpc. In the optical band it shows clear dust lanes and we obtain a total intrinsic column density for the HCR of $1.5 \times 10^{22}\pcmsq$. No blue light in excess of an old stellar population is seen \citep{Farage2010, Canning2011}. It was observed with the  Herschel FIR Observatory from which \cite{Mittal2011} deduced a SFR  of $0.13\Msunpyr$ if it accounts for all the FIR emission. Much lower values of $0.002-0.08\Msunpyr$ were deduced from FUV emission.  The HCF solution requires that the FIR emission be due to the soft X-ray energy absorbed and reradiated by dust in the FIR. Any normal star formation in NGC4696 is then very small (${\rm SFR}\ll 0.05 \times {\rm HCR}$). 
The gas cooled in the HCF becomes invisible, with low-mass star formation being the likely endpoint.

\section{Galaxy formation, young galaxies and cooling flows}

Here we outline a schematic model of galaxy formation. It is based on the original \cite{white1978} model of gas cooling in a dark matter potential well. The new ingredient that  we introduce is a low-mass star formation mode in the high-pressure central region. 

Young galaxies start forming when dense gas falls into dark matter potential wells. The gas is heated to the virial temperature ($T_{\rm V} \sim 10^6-10^7\K$) by shocks and rapidly cools when the radiative cooling time is less than the local freefall time, i.e. $\tau=t_{\rm cool}/t_{\rm freefall}<1$. The cooled gas is turbulent and not necessarily in thermal pressure equilibrium with its surroundings and forms "normal" stars with an Initial Mass Function (IMF) that flattens below $1\Msun$, for example in a   broken-power law \citep{Kroupa2001} or log-normal shape \citep{Chabrier2003}). After a few million years, core-collapse supernovae from massive stars seed the whole region with dust, metals, neutron stars, black holes and turbulent energy. Stellar feedback occurs, churning and expelling gas. Mergers of black holes and accretion of gas build the central black hole. AGN feedback begins. The whole object forms the compact gaseous core of the galaxy, with angular momentum conservation leading to the object flattening into a disc. This phase has  been much discussed, most recently  by \citep{Dekel2023,Silk2024}. 

The next phase, after tens to hundreds of millions of years, is calmer and dominated by gas where the cooling time exceeds the freefall time $\tau>1$, so a slow subsonic inflow sets in, largely  controlled by radiative cooling. This is the cooling flow. The pressure at the centre of nearby objects where the gas finally cools has $nT>10^6\pcmK$, typically $nT\approx10^7\pcmK$, where $n$ and $T$ are the density and temperature of the gas. 

An estimate of $P$ is obtained by equating the bremsstrahlung cooling time ($\sim 2\times 10^{11} T^{1/2}/n)$ to the sound crossing time.
We assume a virialised atmosphere (so sound crossing and freefall times are similar) $\sim r/(10^4 T^{1/2})$,
giving $P=nT = 7\times 10^6 T_6^2/r_2\pcmK$, where the radius $r=100r_2\pc$. If the virial mass exceeds $5\times 10^9\Msun$ then $P_7>1$.

The Jeans mass $M_{\rm J}
=  0.022(1+z)^2/P_7^{1/2}$ \citep{Ferland1994}. 
$M_{\rm J}= 0.022, 0.8$ and $2.7\Msun$ at redshift $z=0, 5, 10$, assuming that the lowest gas temperature is  that of the Cosmic Microwave Background at that redshift.  We propose that stellar objects formed in this phase are principally of low mass,  $M_{\rm J}$ or less, with very few massive stars. Stellar feedback switches off. 

\begin{figure}
	\includegraphics[width=0.47\textwidth]{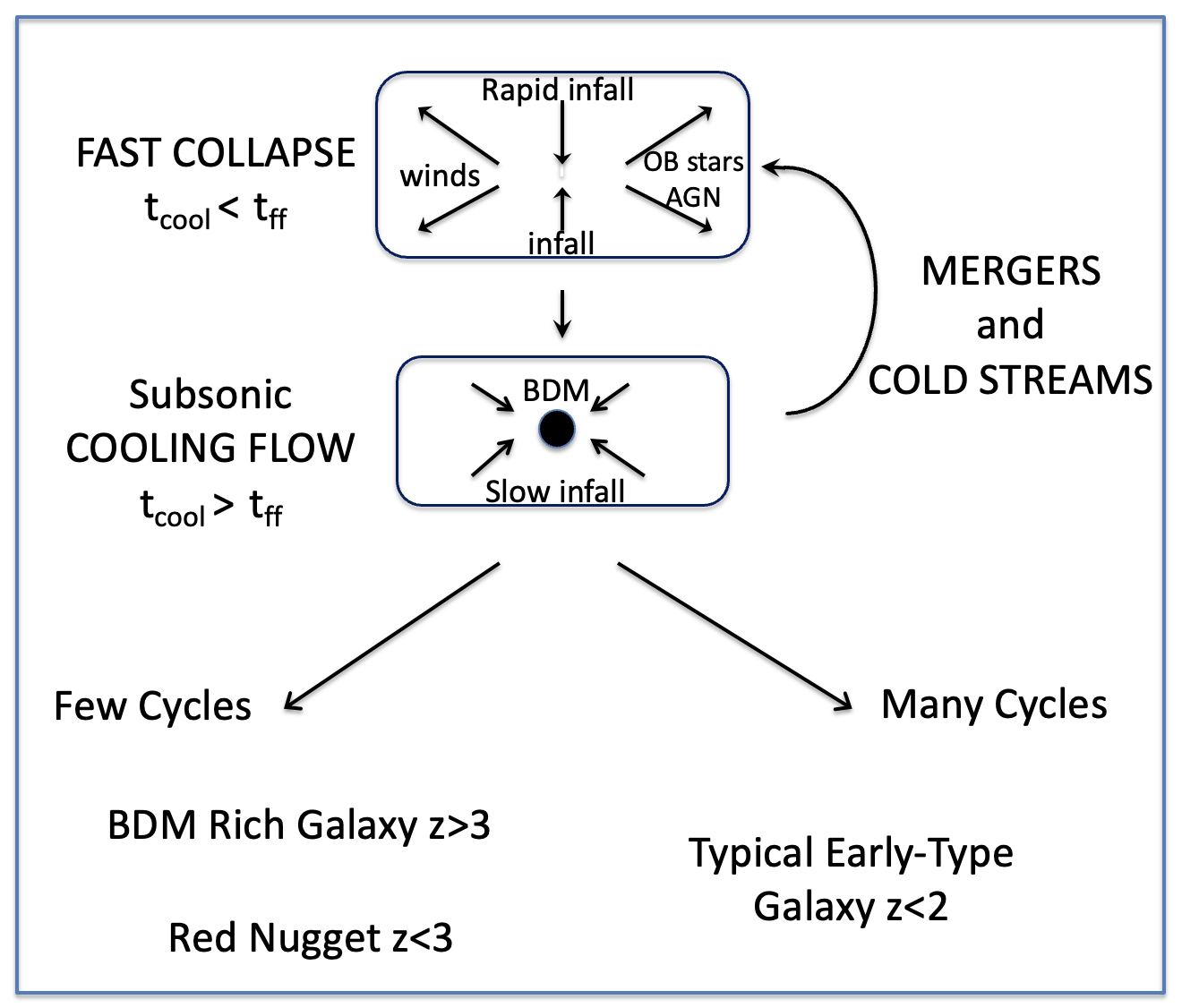}
 \caption{Schematic diagram of the evolution of early-type galaxies envisaged here with time descending downward. The initial fast gravitational collapse and short cooling time leads to rapid infall and normal high mass star formation. This in turn gives winds and stellar feedback. AGN feedback may add to the chaotic state if the black hole is active. When this phase dies down and the cooling time of the interstellar gas becomes greater than the freefall time a subsonic cooling flow occurs. In our picture this leads low-mass star, or baryonic dark matter (BDM) formation. A top-heavy IMF transitions to a bottom-heavy one. Mergers and incoming cold streams create further cycles. Normal Early-Type galaxies at low redshift will have undergone many cycles, but rare objects with few cycles due to either a dense  cluster or empty void environment may be left as BDM rich, red nuggets.  }
\end{figure}

It is not clear that the Jeans mass represents the typical or the maximum final mass of stars formed. If the gas remains approximately isothermal then a collapsing cloud will fragment as the density rises and so the final masses may be much lower than $M_{\rm J}$. Opacity can however provide a lower limit \citep{Rees1976}. We note that opacity can also work on the outside CMB radiation reducing the effect of redshift. We  proceed assuming that $M_{\rm J}$ is more of an {\em upper} limit on the low-mass IMF. The BDM content continues to grow.

The accumulating body of low mass stars orbit the central black hole, transporting angular momentum outward through gravitational instabilities such as the formation of spiral arms and bars, as discussed in HCFIII. This leads to a fraction of them falling inward and being swallowed whole by the black hole once it has grown to about $10^7\Msun$. (it is then too large to tidally disrupt such stars.) Such accretion emits little radiation so the Eddington limit is irrelevant and the massive black hole can grow rapidly, with little AGN feedback acting on the rest of the galaxy.

The galaxy can continue to grow in mass and radius from incoming cold streams of gas or merging with other galaxies, both of which can trigger normal starbursts. Unless the whole hot gaseous atmosphere is disrupted or blown away then  the cooling flow may continue into the centre. Note that all nearby early-type galaxies have extensive hot haloes, with hot gas masses of $10^{10} - 10^{12}\Msun$ \citep{Babyk2018}.

If the galaxy core accretes no further cold gas, either through being in a massive cluster of galaxies e.g NGC1277, or at the opposite extreme being isolated e.g. Mrk1216 (HCFIII), then it can become a rare red nugget (massive compact galaxy) due to a lack of normal star formation. Both NGC1277 and Mrk 1216 lie a factor of 5--10 above the local black-hole mass -- bulge mass relation \citep{walsh2017}.  Studies of nearby nuggets indicate that they have steep low-mass IMFs and ultra-high massive black holes \citep{Ferre015,Ferre2017}, anchoring the above picture at low redshift. 

Very recently, evidence for an overmassive black hole at $z=6.68$ has been reported by \citep{juod2024} based on JWST data. The galaxy, JADES GN 1146115, has a black hole with a mass of $4\times 10^8\Msun$ and  a stellar mass of only $10^9\Msun$ with a SFR of $\sim 1\Msunpyr$, but a dynamical mass of $3 \times 10^9\Msun$. It bears some resemblance to a low redshift nugget, but with a lower total mass.  A low-mass star formation mode as outlined above has the effect of removing gas so suppressing normal star formation whilst enhancing BDM and thus dynamical mass as well as black hole growth. There may be many more objects like this  in the earlier phases of the growth of early-type galaxies. Less than 1 in 1000 massive galaxies at low redshift end up as nuggets, presumably due to the need for special environments, but at high redshift when the number of past mergers or accretion events for a galaxy is small they may be common. We summarise the whole process in Fig. 2.

Direct observational evidence for baryonic dark matter (low-mass stars) in galaxies is difficult to obtain beyond nearby galaxies. Since the work of \citep{vDC2010}  the evidence for a bottom-heavy IMF
at the centre of nearby massive early-type galaxies has however grown significantly \citep{Oldham2018,Smith2020,Mehrgan2019,Mehrgan2024,Parikh2024}. Some work on lensed galaxies at higher redshift also support these findings \citep{Treu2010,Oldham2018b}. A spectacular example of an Einstein Ring around the massive quiescent galaxy JWST-ER1g at $z=2$ \citep{vD2024} enables the total mass of the galaxy within the radius of the ring to be measured with good precision. The authors find that the stellar mass of the galaxy plus its fiducial NFW dark matter component leaves room for a significant BDM component which they ascribe to low-mass stars.  This result is however disputed by \cite{Mercier2023}, who find a more distant redshift for the background galaxy and thus lower mass constraints.

The results in the above examples are consistent with an IMF with a continuous power-law slope to low stellar masses as in the \cite{Salpeter1955} IMF, rather than the break at $0.5\Msun$ in the Kroupa one often used for normal star formation.  We envisage  a separate low-mass star formation mode in the highest pressure regimes rather than a simple extension of normal star formation. Studies and decomposition of the mass and mass to light, $\Upsilon$, profiles of galaxies are clearly very important. 

The central black holes may even be growing now \citep{Farrah2023, Fabian2023b}.  
Numerical simulations are necessary to follow the above scenario in any detail and to make a quantitative comparison of the relative amounts of low-mass stars (BDM), black hole mass, normal Star Formation and hot gas mass in massive galaxies. This is well beyond the scope of the present work.

\section{Conclusions}
Our renewed discovery of cooling flows at the centre of  many massive galaxies, albeit partially hidden by cold absorption, makes the existence of a persistent low-mass star formation mode feasible. It is the most likely repository of the cooled material. The observational evidence for a bottom-heavy IMF in nearby massive galaxies, in which modest cooling flows persist, supports the possibility that this mode has been present during the formation and evolution of many massive galaxies.  It is a factor in all "early" galaxies, from high redshifts to Early-Type Galaxies in the present day. 

A low-mass star formation mode allows for the rapid growth of black holes above $10^7\Msun$, independent of the Eddington limit. it is not then surprising that the most massive black holes in the local Universe lie in deep potential wells hosting hidden cooling flows. Six of the ETGs/BCGs in Fig. 1 host black holes with masses exceeding $10^{10}\Msun$ (Wikipedia.org "List of Most Massive Black Holes")  and another 3 have black holes above $10^9\Msun$.

\section*{Acknowledgements}
We thank Sandro Tacchella, Robert Maiolino and Carolin Crawford for discussions.

\section*{Data Availability}
There are no new data associated with this article.



\bibliographystyle{mnras}
\bibliography{cool_core_Mdot_24} 







\bsp	
\label{lastpage}
\end{document}